\documentclass[aps,pra,showpacs,amsmath,amssymb,superscriptaddress,longbibliography,reprint,10pt]{revtex4-2}
\usepackage{graphicx}
\usepackage{xcolor}
\usepackage{newtxtext,newtxmath}
\usepackage{bbm}
\usepackage{mathrsfs} 
\usepackage{enumerate}
\usepackage[shortlabels]{enumitem}
\usepackage{braket}

\usepackage[colorlinks=true,breaklinks=true,allcolors=blue]{hyperref}

\definecolor{michael}{rgb}{.2,.5,.6}

\definecolor{cg}{rgb}{1,0,0}

\definecolor{philip}{rgb}{.2,.2,.7}

\newcommand\id{{\mathbbm{1}}}

\DeclareMathOperator{\Tr}{{Tr}}
\DeclareMathOperator{\myIm}{\mathrm{Im}}
\DeclareMathOperator{\myRe}{\mathrm{Re}}

\newcommand\articletype{Letter }

\makeatletter
\renewcommand{\p@paragraph}{}
\makeatother

\makeatletter
\let\cat@comma@active\@empty
\makeatother

\makeatletter
\g@addto@macro\bfseries{\boldmath}
\makeatother

\begin{document}

\title{Ancilla-Free Measurement of Out-of-Time-Ordered Correlation Functions:\texorpdfstring{\\}{} General Measurement Protocol and Rydberg Atom Implementation}  

\author{Michael Kastner}
\affiliation{Institute of Theoretical Physics, University of Stellenbosch, Stellenbosch 7600, South Africa}
\affiliation{Hanse-Wissenschaftskolleg, Lehmkuhlenbusch 4, 27753 Delmenhorst, Germany}

\author{Philip Osterholz}

\author{Christian Gross}
\affiliation{Physikalisches Institut, Eberhard Karls Universität Tübingen, 72076 Tübingen, Germany}

\date{\today}

\begin{abstract}
We introduce a protocol that gives access to out-of-time-ordered correlation functions in many-body quantum systems. Unlike other such protocols, our proposal, which can be applied to arbitrary initial states, neither requires ancilla degrees of freedom to the quantum system of interest, nor has the need for randomized measurements. Nontrivial experimental capabilities required to implement the protocol are single-site measurements, single-site rotations, and backwards time evolution. To exemplify the implementation of the protocol, we put forward a strategy for Hamiltonian sign inversion $H\to-H$ in arrays of Rydberg-dressed atoms. In this way, a complete and practical toolbox is obtained for the measurement of out-of-time-ordered correlations in equilibrium and nonequilibrium situations.
\end{abstract}


\maketitle 

``Scrambling'' refers to the process where, under unitary time evolution of a many-body quantum system, initially local information disperses into many-body entanglement and spreads over increasingly larger regions of the system. In this process, information, while in principle conserved, becomes inaccessible to local measurements, resulting in an effective loss of memory. Quantum information scrambling rose to prominence in the context of the black hole information problem \cite{SekinoSusskind08,*ShenkerStanford14}, and accompanies the dynamics of thermalization in isolated quantum systems \cite{Deutsch91,*Srednicki94,*GogolinEisert16}.

A strategy to quantify the size of the region across which quantum information is spread under the system dynamics consists in considering two operators $V$ and $W$ that are initially supported on separated regions of space, implying $[W,V]=0$. The commutator $C(t)=\left\langle\lvert[W(t),V]\rvert^2\right\rangle$ then quantifies the degree to which the time-evolved operator $W(t)=\exp(iHt)W\exp(-iHt)$ spreads into the support of $V$, indicating whether or not the region over which quantum information can be scrambled in time $t$ extends into the support of $V$. Moreover, the semiclassical limit of $C$ quantifies the sensitivity of the dynamics to small changes in the initial conditions, which suggests to interpret $C$ as a measure of quantum chaoticity \cite{MaldacenaShenkerStanford16}. Closely related to $C$, and often more accessible, is the out-of-time-ordered correlation (OTOC) \footnote{For unitary operators $V$ and $W$ satisfying $[W,V]=0$, a short calculation yields $\myRe[F(t)]=1-\left\langle\lvert[W(t),V]\rvert^2\right\rangle/2$.}
\begin{equation}\label{e:OTOCgeneral}
F(t) = \Braket{W^\dagger(t)V^\dagger W(t)V}.
\end{equation}
Such correlation functions have been used in early studies of electron scattering off impurities in models of superconductors \cite{LarkinOvchinnikov69}, are related to the Renyi entropy and the generation of entanglement \cite{Hosur_etal16,Fan_etal17}, and have been employed to distinguish many-body localized phases from thermalizing ones \cite{Fan_etal17,Huang_etal17,*Chen_etal17}.

\begin{figure}\centering      
   \includegraphics{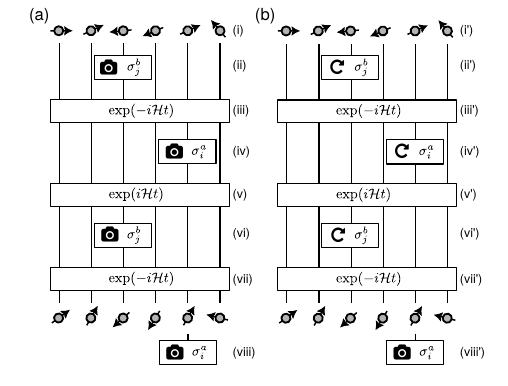}      
 \caption{Sequence of steps for accessing the real part (a) and imaginary part (b) of an OTOC. From top to bottom: (i) initial state, (ii) projective measurement or spin rotation at site $j$, (iii) forward time evolution under the Hamiltonian $H$, (iv) projective measurement or spin rotation at site $i$, (v) backward time evolution under $H$, (vi) projective measurement or spin rotation at site $j$, (vii) forward time evolution under $H$, (viii) projective measurement at site $i$.}
 \label{f:cartoon}
\end{figure}

A setting of particular physical interest is when $V$ and $W$ are local Hermitian operators that correspond to experimentally measurable observables. But even in such a, presumably ``simpler,'' situation, $F$ in Eq.\ \eqref{e:OTOCgeneral} is the expectation value of a, in general, non-Hermitian and nonlocal operator. A direct measurement of such an expectation value is unrealistic in all but the most trivial special cases, and one therefore has to resort to indirect measurement protocols in order to gain access to OTOCs. Such measurement protocols for OTOCs essentially interpret the product of operators on the right-hand side of Eq.\ \eqref{e:OTOCgeneral} as a sequence of operations. Because of the lack of time-ordering that is characteristic for OTOCs, this sequence of operations involves forward-in-time as well as backward-in-time evolving unitaries, which, while challenging in general, are available in certain experimental platforms such as atoms in optical lattices~\cite{braun2013} and in all-to-all interacting neutral atom or ion systems~\cite{linnemann2016, gilmore2021a, colombo2022}. While most of the proposed measurement protocols for OTOCs require backward-in-time evolution, other strategies, each with their own merits and challenges, are known as well \cite{Vermersch_etal19,*Blocher_etal22}. Some aspects of OTOCs have already been measured in small systems~\cite{li2017b, wei2018, landsman2019, nie2020, chen2020, joshi2020}.

Here we will focus on measurement protocols for OTOCs that make use of backwards time evolution. Several such protocols have been proposed by now \cite{Swingle_etal16,*GarciaAlvarez_etal17,Sundar_etal22}, some of which have been implemented experimentally \cite{Gaerttner_etal17,*Li_etal17,*Braumueller_etal22,Mi_etal21}. While none of these protocols is easy to implement, some are less challenging, but at the expense of being applicable only to specific choices of $V$ and/or specific initial states \cite{Gaerttner_etal17,*Li_etal17,*Braumueller_etal22,Sundar_etal22}. The protocol put forward in Ref.~\cite{Swingle_etal16,*GarciaAlvarez_etal17}, on the other side, is applicable to arbitrary $V$, $W$ and arbitrary initial states. It is, however, significantly more difficult to implement, as it requires, in addition to backwards time evolution, the capability to couple an ancilla qubit to the system of interest and to create entanglement between the two. An experimental realization of this ancilla-based protocol was reported in Ref.\ \cite{Mi_etal21} for a two-dimensional array of superconducting qubits, which is known for the exquisite level of experimental controllability. In many other experimental platforms, however, the simultaneous requirements of backwards time evolution and ancilla coupling poses an obstacle. This motivates the search for alternative measurement protocols, ideally of broad applicability and with more moderate experimental requirements.

The first main result of this \articletype is a measurement protocol for OTOCs that requires neither ancilla quantum degrees of freedom to be coupled to the system, nor interferometric techniques, nor averaging over randomized initial states. In addition to backwards time evolution, the main experimental requirement to execute the protocol are single-site measurements and single-site rotations in a system of qubits or spin-$1/2$ degrees of freedom; see Fig.~\ref{f:cartoon} for an illustration. Single-site resolution and addressability are key capabilities of quantum computing platforms and are readily available in a variety of experimental settings. 

As the second main result of this \articletype we propose a technique, based on microwave-assisted Rydberg dressing~\cite{sevincli2014, petrosyan2014, bijnen2015}, that facilitates Hamiltonian sign inversion, and hence backwards time evolution, in array of ultracold Rydberg-dressed atoms. Combining this technique with local in-sequence readout and local rotation techniques, our measurement protocol becomes a full-fledged tool for the measurement of OTOCs in a relevant experimental platform, while avoiding the challenge of having to couple ancillas to individual sites. Since the protocol is not restricted to specific initial states, OTOCs can be measured and analyzed in arbitrary equilibrium and nonequilibrium situations. The sign-inversion technique is interesting also in the context of other applications, for example, for quantum simulation at finite energies~\cite{lu2021} or quantum metrology~\cite{davis2016,*frowis2016,*macri2016}.

{\em OTOC measurement protocol.---}We consider an arbitrary network of $N$ qubits, including regular lattices as special cases. 
Time evolution is assumed to be unitary, generated by a time-independent Hamiltonian $H$, but is arbitrary otherwise, allowing for multi-site interactions (beyond pair interactions) as well as interactions of arbitrarily long range.

Our main technical requirement is that the operators $V$ and $W$ in Eq.\ \eqref{e:OTOCgeneral} have only two distinct (albeit possibly degenerate) eigenvalues. In the context of qubit systems, this is a very natural and not particularly restrictive setting. For notational simplicity we will in the following choose single-site Pauli spin operators as observables, $W=\sigma_i^a$ and $V=\sigma_j^b$, where $i$ and $j$ denote sites on the network and $a,b\in\{x,y,z\}$ label spin components. The object of study is the OTOC
\begin{equation}\label{e:OTOC}
C(t):=\Tr\left[\rho\sigma_i^a(t)\sigma_j^b\sigma_i^a(t)\sigma_j^b\right],
\end{equation}
where $\rho$ is the initial density operator. Note that, even though each factor of the operator product on the right-hand side of Eq.\ \eqref{e:OTOC} is Hermitian, the product of operators in general is not, and hence $C$ can be complex.

{\em Measurement protocol for the real part.---}We show that the real part of $C$ can be obtained by interpreting the operator product inside the trace of Eq.\ \eqref{e:OTOC} as {\em measurements}\/ of the occurring spin operators, interspersed with time evolutions \footnote{This strategy has been employed in a different context in Ref.~\cite{Uhrich_etal17}.}. This is by no means a trivial statement, as Eq.\ \eqref{e:OTOC} describes unitary evolution, and measurements are known to disturb unitary evolution due to wave function collapse. However, when probing bivariate observables, these disturbing effects, which do occur, cancel out exactly when using the following measurement protocol, illustrated in Fig.~\ref{f:cartoon}:
\begin{enumerate}[start=1,label={(\roman*)}]
\setlength{\itemsep}{0pt}
\setlength{\parskip}{0pt}
\item Prepare the initial state $\ket{\psi}$.\label{i:first}
\item Projectively measure the observable $\sigma_j^b$ and record the outcome ($+$ or $-$).\label{i:meas1}
\item Time-evolve unitarily until time $t$.
\item Projectively measure the observable $\sigma_i^a$ and record the outcome.\label{i:meas2}
\item Evolve {\em backwards}\/ in time for a time $t$.
\item Projectively measure the observable $\sigma_j^b$ and record the outcome.\label{i:meas3}
\item Time-evolve unitarily until time $t$.
\item Projectively measure the observable $\sigma_i^a$ and record the outcome.\label{i:last}
\item Repeat \ref{i:first}--\ref{i:last} many times and record the relative frequencies of the combinations of measurement outcomes $(+++\,+)$, $(+++\,-)$, $(++-\,+)$, etc., occurring in each of the measurement sequences.
\item Use these relative frequencies to estimate the corresponding probabilities $P_{++++}$, $P_{+++-}$, $P_{++-+}$, etc., and calculate the correlation function
\begin{equation}\label{e:CorrProj}
\mathscr{C}(t):=\sum_{o_1,o_2,o_3,o_4\in\{-1,+1\}}o_1 o_2 o_3 o_4 P_{o_1 o_2 o_3 o_4}.
\end{equation}
\end{enumerate}
We show in Appendix \ref{s:proof_proj} that
\begin{equation}\label{e:CeqReC}
2\mathscr{C}(t)-1=\myRe C(t),
\end{equation}
i.e., the real part of the desired OTOC \eqref{e:OTOC} is obtained by applying the above protocol that ``naively'' disregards the effect of measurement backaction. The key experimental capabilities to execute the protocol are unitary backwards time evolution and projective measurements of single qubits.

\begin{figure}\centering
\includegraphics[width=0.9\linewidth]
{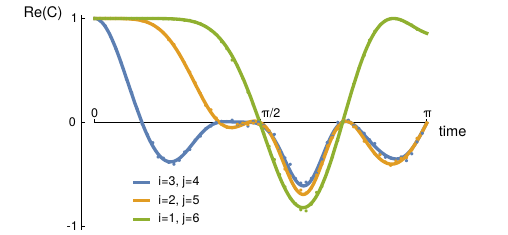}
\caption{\label{f:examples} 
Real part of the OTOC $C$ as a function of time $t$ in an $XY$ spin chain of length $N$. Different colors correspond to different pairs of lattice sites in definition \eqref{e:OTOC}, as specified in the legend. Solid lines show exact results for $C$, the dots mark estimated values obtained by calculating $2\mathscr{C}-1$ from samples of size $10^4$. The further apart the sites $i$ and $j$, the longer it takes until the OTOC starts to deviate from its initial value of $1$, which is a manifestation of quasilocality in a system with short-range interactions.}
\end{figure}

\begin{figure}\centering
\includegraphics{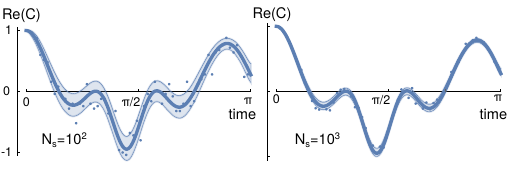}
\caption{\label{f:errors} 
Magnitude of the statistical fluctuations in the estimator $2\mathscr{C}-1$, obtained from Eq.\ \eqref{e:CorrProj} and using samples of sizes $N_s=10^2$ (left) and $N_s=10^3$ (right). All data are for the OTOC \eqref{e:OTOC} with $i=2$ and $j=3$ on a chain of $N=4$ sites.
Solid lines show the exact result, the dots are estimated values, fluctuating around the exact value. The shaded area around the exact result indicates an estimate of the statistical error, obtained by drawing multiple samples, each of size $N_s$, from the probability distribution of $P_{\pm\pm\pm\pm}$ and then calculate the standard deviation of the estimated values.
}
\end{figure}

{\em Unitarily evolved vs.\ projectively measured OTOCs.---}Equation \eqref{e:CeqReC} is an exact relation between the OTOC $C$ and the correlation function $\mathscr{C}$, where the latter is determined by the probabilities $P_{\pm\pm\pm\pm}$. In an experimental realization of the measurement protocol, these probabilities must be estimated through finite sample averages. The sampling introduces statistical errors, which, by error propagation, cause errors in $C$. We illustrate the magnitude of these errors, and hence the performance of the proposed measurement protocol when constrained by limited resources, for spin chains with nearest-neighbor $XY$ couplings,
\begin{equation}
H=-\sum_{k=1}^{N-1}\left(\sigma_k^x \sigma_{k+1}^x + \sigma_k^y \sigma_{k+1}^y\right).
\end{equation}
For this Hamiltonian and using the initial state $\rho=(\ket{\uparrow\cdots\uparrow}\bra{\uparrow\cdots\uparrow})^{\otimes N}$, we calculate the probabilities $P_{\pm\pm\pm\pm}$ as given in Appendix \ref{s:proof_proj}. To simulate a finite number of experimental runs of the measurement protocol, we draw pseudorandom numbers according to these probabilities and calculate a finite-sample estimator of $\mathscr{C}$. In Fig.\ \ref{f:examples}, the real part $\myRe C$ of the exact OTOC is compared to the finite-sample estimator of $2\mathscr{C}-1$ for samples of size $N_s=10^4$. While the agreement with the exact result is excellent, small statistical errors are visible in the estimator. To assess the magnitude of the statistical errors, we compare the estimators obtained with sample sizes $N_s=10^2$, and $10^3$ in Fig.\ \ref{f:errors}. Statistical errors are found to decrease quickly with increasing sample size, and a moderate value of $N_s=10^3$ is sufficient to obtain relative statistical errors of only a few percent.

{\em Measurement protocol for the imaginary part.---}The imaginary part of an OTOC is known to contain information that is complementary to that of the real part \cite{Sajjan_etal23}, and obtaining both is therefore desirable. The corresponding measurement protocol we present here requires, in addition to backward time evolution, the experimental capability to perform single-qubit rotations at sites $i$ and $j$. The imaginary part of the OTOC \eqref{e:OTOC} can be obtained by the following sequence of operations (see Fig.~\ref{f:cartoon} for an illustration):
\begin{enumerate}[start=1,label={(\roman*')}]
\setlength{\itemsep}{0pt}
\setlength{\parskip}{0pt}
\item Prepare the initial state $\ket{\psi}$.\label{i:firstrot}
\item Rotate spin $j$ by an angle $\theta_1$ around the $b$-direction.\label{i:rot1}
\item Time-evolve unitarily until time $t$.
\item Rotate spin $i$ by an angle $\theta_2$ around the $a$-direction.\label{i:rot2}
\item Evolve {\em backwards}\/ in time for a time $t$.
\item Rotate spin $j$ by an angle $\theta_3$ around the $b$-direction.\label{i:rot3}
\item Time-evolve unitarily until time $t$.\label{i:lasttime}
\item Projectively measure the observable $\sigma_i^a$.\label{i:lastrot}
\item Repeat \ref{i:firstrot}--\ref{i:lastrot} many times and record the empirical mean of the measurement outcome.\label{i:final}
\end{enumerate}
This empirical mean gives an estimator of the expectation value $\braket{\sigma_i^a}_{t,\theta_1,\theta_2,\theta_3}$ with respect to the state at the end of step \ref{i:lasttime} of the above protocol. We show in Appendix \ref{s:proof_rot} that
\begin{multline}\label{e:CeqImC}
\bigl\langle\sigma_i^a\bigr\rangle_{t,-\theta_1,-\theta_2,-\theta_3}-\bigl\langle\sigma_i^a\bigr\rangle_{t,\theta_1,\theta_2,\theta_3}\\
-\bigl\langle\sigma_i^a\bigr\rangle_{t,-\theta_1,\theta_2,-\theta_3}+\bigl\langle\sigma_i^a\bigr\rangle_{t,\theta_1,-\theta_2,\theta_3}\\
= 4\sin(\theta_2) \sin(\theta_1+\theta_3/2) \sin(\theta_3/2) \myIm C(t).
\end{multline}
Hence, the imaginary part of the OTOC \eqref{e:OTOC} is obtained by performing the above protocol \ref{i:firstrot}--\ref{i:lastrot} for the four sets of rotation angles $(-\theta_1,-\theta_2,-\theta_3)$, $(\theta_1,\theta_2,\theta_3)$, $(-\theta_1,\theta_2,-\theta_3)$, and $(\theta_1,-\theta_2,\theta_3)$. A simple and, in some sense, optimal choice is $\theta_1=\theta_2=\theta_3=\pi/2$, in which case the right-hand side of Eq.~\eqref{e:CeqImC} simplifies to $2\myIm C(t)$.

{\em Experimental implementation with Rydberg atom arrays.---}The requirements to implement the above described protocols are coherent forward and backward time evolution under an effective spin-$1/2$ Hamiltonian, as well as the capability to conduct single-site projective measurements and single-site spin rotations. Among the experimental platforms that allow for coherent dynamics of many-body qubit systems, arrays of ultracold Rydberg atoms in optical tweezers are particularly suitable (see Ref.~\cite{browaeys2020} for a review). Optical tweezers allow the experimenter to arrange and hold the atoms in a lattice geometry of choice, with lattice constants of the order of micrometers. To make the atoms interact over such distances, they are excited to Rydberg states, i.e., atomic states with a large principal quantum number. Various schemes to emulate spin-$1/2$ Hamiltonians in such arrays have been devised, giving rise to coherent spin dynamics under Ising- or $XY$-type Hamiltonians. Moreover, single-site control is well established for these platforms~\cite{graham2022a, bluvstein2023}. In the following we present a solution to the remaining challenge for a successful implementation of our OTOC measurement protocol in Rydberg platforms, namely backwards time evolution or, equivalently, sign inversion of the Hamiltonian.

{\em Time inversion in Rydberg atom arrays.---}The experimental scheme we propose is based on microwave-assisted Rydberg dressing~\cite{sevincli2014,*petrosyan2014,*bijnen2015}. Rydberg dressing naturally leads to spin-1/2 degrees of freedom being encoded in two low-lying and long-lived atomic states, with long-range Ising interactions between the spins~\cite{gil2014}. The strength of these interactions can be dynamically tuned by choosing the detuning of the light field, which off-resonantly couples one of the states to a Rydberg state. Several experiments demonstrated this technique in optical lattices and tweezers~\cite{jau2016,*zeiher2016,*zeiher2017a,*guardado-sanchez2021,*hollerith2022,*steinert2023,*eckner2023}. The sign of the induced Ising interaction is determined by the sign of the interaction in the laser-addressed Rydberg state. Implementing time reversal with this technique is not straightforward, as it requires the existence of Rydberg states with opposite sign and near-equal magnitude of the interaction potential. Even if a specific pair of Rydberg states is identified that meets this special condition, interaction reversal is still a technically demanding task requiring two distinct laser frequencies.

\begin{figure}[t]
   \centering      
   \includegraphics{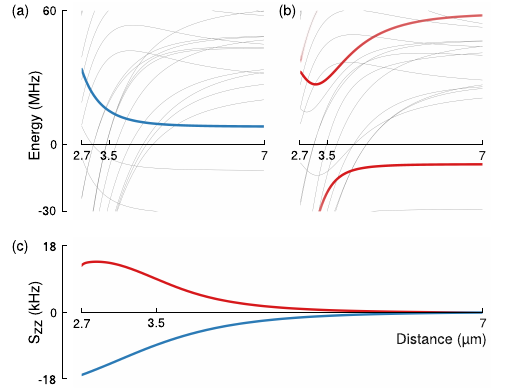}      
 \caption{\label{f:dressing}Illustration of microwave-assisted potential inversion for the example of $^{39}$K. (a)~Interaction potential between pairs of atoms for various eigenstates of the dipolar interaction. The blue line highlights the pair potential between atoms in the \(\ket{50S_{1/2}, m_F=1/2}\) state. The energy corresponding to the laser frequency is chosen as the energy-zero level and the polarization is such that only the highlighted level is coupled from the groundstate. (b)~Switching on microwave radiation with a frequency of 27.592 GHz, the \(\ket{50S_{1/2}, m_J=1/2}\) state is coupled to the \(\ket{50P_{1/2}, m_J=1/2}\) state. As a result, the pair potentials highlighted in red emerge from the blue line in (a) as microwave-dressed pair-states. The plotted data assumes linear polarization and a realistic choice of the Rabi frequency of about $30\,$MHz. The shading of the colored lines in (a) and (b) is chosen proportional to the laser coupling strength. (c)~Resulting dressed Ising interactions in the electronic ground state for a laser red-detuned by 4 MHz with respect to the single-atom resonance and 4.4 MHz blue-detuned to the lower of the microwave-shifted states. The laser Rabi frequency has been set to 2 MHz.
}
\end{figure}

Here we show, for the example of $^{39}$K atoms, that this difficulty can be overcome, and sign inversion of the Ising interactions can be implemented, by strong microwave coupling of the van der Waals pair-potential curve to a dipolar potential curve. The blue line in Fig.~\ref{f:dressing}(a) shows the interaction potential between pairs of atoms in the \(\ket{50S_{1/2}, m_F=1/2}\) state, where the zero-energy level is chosen to coincide with the laser with the laser-targeted energy. When switching on microwave radiation with a frequency of 27.592 GHz, the \(\ket{50S_{1/2}, m_J=1/2}\) state is coupled to the \(\ket{50P_{1/2}, m_J=1/2}\) state, resulting in the pair potential shown in red in Fig.~\ref{f:dressing}(b). The splitting of the two emerging potential branches at large distances is determined by the Autler-Townes splitting of the single-atom state. At shorter distances of around 3-4 $\mu$m, the signature of an avoided level crossing is visible. For microwave interaction of a suitably chosen strength, the avoided level crossing is opened up to an extent such that the lower branch in Fig.~\ref{f:dressing}(b) becomes the reflection image of the blue line in Fig.~\ref{f:dressing}(a) around the laser-targeted energy. This property of one pair potential being the reflection image of another then gets imprinted onto the corresponding Ising interactions, as illustrated in Fig.~\ref{f:dressing}(c). These results were obtained with the ``pairinteraction'' software~\cite{weber2017}, which diagonalizes the two-atom Hamiltonian, including magnetic fields and dipolar interactions. We extended the software to also include the effect of microwave coupling in the diagonalization. Figure ~\ref{f:dressing}(c) illustrates that, by merely switching on or off the microwave-induced coupling, the sign of the Ising interaction is inverted, and hence backward time evolution can be realized. The Rydberg dressing approach discussed here has the advantageous feature that the atoms remain trapped in the optical tweezers during the entire sequence, and that forces between the atoms remain small. This avoids decoherence due to uncontrolled coupling to motional degrees of freedom, which would corrupt time reversal.


{\em Summary and outlook.---}The results presented in this \articletype lay out a path towards measuring OTOCs for arbitrary many-body qubit systems and initial states. Unlike other measurement protocols of that generality, our proposal requires neither randomized measurements nor the use of ancilla degrees of freedom. In our protocol, real and imaginary parts of the OTOC are measured separately, both requiring local control (single-site measurements and single-site rotations, respectively) in the experiment. In addition, coherent forward- as well as backward-in-time evolution are needed.

We demonstrate the feasibility of the proposed measurement protocol for Rydberg-dressed atoms in optical tweezers. While single-site measurements and rotations are readily available in this platform, Hamiltonian sign inversion, which facilitates backward-in-time evolution, has hitherto been missing. We introduced a method that facilitates sign inversion through microwave-assisted Rydberg dressing, thus completing the toolset for successfully implementing our OTOC measurement protocol in arrays of Rydberg atoms. This paves the way for experimental explorations of equilibrium and nonequilibrium situations through OTOCs in these platforms, including the detection of quantum chaos~\cite{GarciaMata_etal23,*GonzalezAlonso_etal}, the monitoring of thermalization and the scrambling of quantum information~\cite{Fan_etal17,Huang_etal17,*Chen_etal17,Sajjan_etal23}, or more exotic tasks like the probing of excited-state quantum phase transitions~\cite{WangPerezBernal19}. We expect our protocol to become applicable in other many-body qubit systems or analog quantum simulators of spin-$1/2$ Hamiltonians in the near future. 

{\em Note added:} Upon completion of this \articletype we became aware of Ref.~\cite{Geier_etal}, in which sign inversion in a bulk Rydberg gas implementing the dipolar XXZ model was demonstrated.

\acknowledgments
We acknowledge funding from the Deutsche Forschungsgemeinschaft via projects GR\,4741/4-1, SPP 1929 GiRyd (GR\,4741/5-1) and FOR 5413 (GR\,4741/6-1), a Heisenberg professorship (GR\,4741/3-1 and GR\,4741/7-1) and from the Alfried Krupp von Bohlen und Halbach foundation.

\appendix

\section{Proof of Eq.~\texorpdfstring{\eqref{e:CeqReC}}{(4)}}
\label{s:proof_proj}
Denote by $\Pi_i^{\pm_a}$ the projector onto the $\pm$-eigenstate of the Pauli operator $\sigma_i^a$.

In the course of the protocol \ref{i:first}--\ref{i:last} the system state goes through the following stages:
\begin{itemize}
\setlength{\itemsep}{0mm}
\item[\ref{i:first}] Initial state $\rho$.
\item[\ref{i:meas1}] After the first measurement, and depending on the outcome $\pm$ of the measurement, the state is
\begin{equation}\label{e:psi1}
\rho_1^\pm=\Pi_j^{\pm_b}\rho\Pi_j^{\pm_b}/P_\pm,
\end{equation}
where
\begin{equation}
P_{\pm}=\Tr\left(\rho\Pi_j^{\pm_b}\right)
\end{equation}
is the probability of measuring $+$ or $-$.
\item[\ref{i:meas2}] After the second measurement, and depending on the outcome $\pm$ of the first measurement \ref{i:meas1} and the outcome $\pm$ of the second measurement \ref{i:meas2}, the state is
\begin{equation}
\rho_2^{\pm\pm}=\Pi_i^{\pm_a}e^{-iHt}\rho_1^\pm e^{iHt}\Pi_i^{\pm_a}/P_{\pm|\pm},
\end{equation}
where
\begin{equation}
P_{\pm|\pm}=\Tr\left(e^{-iHt}\rho_1^\pm e^{iHt}\Pi_i^{\pm_a}\right)
\end{equation}
is the conditional probability of measuring $\pm$ in the second measurement after having measured $\pm$ in the first measurement.
\item[\ref{i:meas3}] Similarly one obtains 
\begin{equation}
\rho_3^{\pm\pm\pm}=\Pi_j^{\pm_b}e^{iHt}\rho_2^{\pm\pm}e^{-iHt}\Pi_j^{\pm_b}/P_{\pm|\pm\pm}
\end{equation}
with
\begin{equation}
P_{\pm|\pm\pm}=\Tr\left(e^{iHt}\rho_2^{\pm\pm}e^{-iHt}\Pi_j^{\pm_b}\right)
\end{equation}
after the third measurement.
\item[\ref{i:last}] And
\begin{equation}
\rho_4^{\pm\pm\pm\pm}=\Pi_i^{\pm_a}e^{-iHt}\rho_3^{\pm\pm\pm}e^{iHt}\Pi_i^{\pm_a}/P_{\pm|\pm\pm\pm}
\end{equation}
with
\begin{equation}
P_{\pm|\pm\pm\pm}=\Tr\left(e^{-iHt}\rho_3^{\pm\pm\pm}e^{iHt}\Pi_i^{\pm_a}\right)
\end{equation}
after the fourth measurement.
\end{itemize}
The probability of finding a specific sequence of the four measurement outcomes $\pm$ is then given by
\begin{equation}\label{e:probabilities}
P_{\pm\pm\pm\pm}=P_\pm P_{\pm|\pm} P_{\pm|\pm\pm} P_{\pm|\pm\pm\pm}.
\end{equation}
Inserting Eqs.\ \eqref{e:psi1}--\eqref{e:probabilities} into the correlation function \eqref{e:CorrProj}, the latter can be written as
\begin{multline}
\mathscr{C}(t)=\sum_{o_1,o_2,o_3,o_4\in\{-,+\}}o_1 o_2 o_3 o_4\\
\times\Tr\left[\Pi_i^{o_4}(t)\Pi_j^{o_3}\Pi_i^{o_2}(t)\Pi_j^{o_1}\rho\Pi_j^{o_1}\Pi_i^{o_2}(t)\Pi_j^{o_3}\right],
\end{multline}
where $\Pi_i^o(t):=e^{iHt}\Pi_i^oe^{-iHt}$. Making use of the spectral representation
\begin{equation}
\sigma_i^a=\Pi_i^{+_a} - \Pi_i^{-_a}=\sum_{o_4\in\{-,+\}}o_4\Pi_i^{o_4}
\end{equation}
as well as the linearity of the trace, one obtains
\begin{multline}\label{e:Ccurl2}
\mathscr{C}(t)=\sum_{o_1,o_2,o_3\in\{-,+\}}o_1 o_2 o_3\\
\times\Tr\left[\sigma_i^a(t)\Pi_j^{o_3}\Pi_i^{o_2}(t)\Pi_j^{o_1}\rho\Pi_j^{o_1}\Pi_i^{o_2}(t)\Pi_j^{o_3}\right].
\end{multline}
Using the completeness relation $\Pi_j^{o_3}=\id_j-\Pi_j^{-o_3}$ on the second occurrence of $\Pi_j^{o_3}$ in \eqref{e:Ccurl2}, $\mathscr{C}$ can be rewritten as
\begin{multline}\label{e:Ccurl3a}
\mathscr{C}(t)=\sum_{o_1,o_2\in\{-,+\}}o_1 o_2\\
\times\Tr\left[\sigma_i^a(t)\left(\sum_{o_3\in\{-,+\}}o_3\Pi_j^{o_3}\right)\Pi_i^{o_2}(t)\Pi_j^{o_1}\rho\Pi_j^{o_1}\Pi_i^{o_2}(t)\right]\\
-\sum_{o_1,o_2,o_3\in\{-,+\}}o_1 o_2 o_3\\
\times\Tr\left[\sigma_i^a(t)\Pi_j^{o_3}\Pi_i^{o_2}(t)\Pi_j^{o_1}\rho\Pi_j^{o_1}\Pi_i^{o_2}(t)\Pi_j^{-o_3}\right].
\end{multline}
The second line of \eqref{e:Ccurl3a} can be simplified by recognizing that the term in round brackets is the spectral representation of $\sigma_j^b=\Pi_j^{+_b} - \Pi_j^{-_b}$. Writing out the remaining $o_3$-sum in the bottom two lines of \eqref{e:Ccurl3a} and using the cyclic invariance of the trace, one obtains
\begin{multline}\label{e:Ccurl3c}
\mathscr{C}(t)=\sum_{o_1,o_2\in\{-,+\}}o_1 o_2\\
\times\biggl(\Tr\left[\sigma_i^a(t)\sigma_j^b\Pi_i^{o_2}(t)\Pi_j^{o_1}\rho\Pi_j^{o_1}\Pi_i^{o_2}(t)\right]\\
-\Tr\left[\sigma_i^a(t)\Pi_j^{+_b}\Pi_i^{o_2}(t)\Pi_j^{o_1}\rho\Pi_j^{o_1}\Pi_i^{o_2}(t)\Pi_j^{-_b}\right]\\
+\Tr\left[\Pi_j^{-_b}\Pi_i^{o_2}(t)\Pi_j^{o_1}\rho\Pi_j^{o_1}\Pi_i^{o_2}(t)\Pi_j^{+_b}\sigma_i^a(t)\right]\biggr).
\end{multline}
The operator products in the second and third trace in Eq.\ \eqref{e:Ccurl3c} are Hermitian conjugates of each other. Hence, the difference of the traces can be expressed as an imaginary part, yielding
\begin{multline}\label{e:Ccurl3d}
\mathscr{C}(t)=2i\myIm c_3+\sum_{o_1,o_2\in\{-,+\}}o_1 o_2\\
\times\Tr\left[\sigma_i^a(t)\sigma_j^b\Pi_i^{o_2}(t)\Pi_j^{o_1}\rho\Pi_j^{o_1}\Pi_i^{o_2}(t)\right]
\end{multline}
with
\begin{multline}\label{e:c3}
c_3=\sum_{o_1,o_2\in\{-,+\}}o_1 o_2\\
\times\Tr\left[\Pi_j^{-_b}\Pi_i^{o_2}(t)\Pi_j^{o_1}\rho\Pi_j^{o_1}\Pi_i^{o_2}(t)\Pi_j^{+_b}\sigma_i^a(t)\right].
\end{multline}

Next we apply the completeness relation $\Pi_i^{o_2}=\id_j-\Pi_i^{-o_2}$ and the corresponding spectral representation to the first occurrence of $\Pi_i^{o_2}$ in Eq.\ \eqref{e:Ccurl3d}, which yields
\begin{multline}\label{e:Ccurl4a}
\mathscr{C}(t)=2i\myIm c_3+\sum_{o_1\in\{-,+\}}o_1\Tr\left[\sigma_j^b\Pi_j^{o_1}\rho\Pi_j^{o_1}\right]\\
-\sum_{o_1,o_2\in\{-,+\}}o_1 o_2
\Tr\left[\sigma_i^a(t)\sigma_j^b\Pi_i^{-o_2}(t)\Pi_j^{o_1}\rho\Pi_j^{o_1}\Pi_i^{o_2}(t)\right].
\end{multline}
Applying the same completeness relation and spectral representation to the second occurrence of $\Pi_i^{o_2}$ in Eq.\ \eqref{e:Ccurl3d}, one obtains
\begin{multline}\label{e:Ccurl4b}
\mathscr{C}(t)=2i\myIm c_3+\sum_{o_1\in\{-,+\}}o_1\Tr\left[\sigma_i^a(t)\sigma_j^b\sigma_i^a(t)\Pi_j^{o_1}\rho\Pi_j^{o_1}\right]\\
+\sum_{o_1,o_2\in\{-,+\}}o_1 o_2
\Tr\left[\sigma_i^a(t)\sigma_j^b\Pi_i^{-o_2}(t)\Pi_j^{o_1}\rho\Pi_j^{o_1}\Pi_i^{o_2}(t)\right].
\end{multline}
Summing Eqs.\ \eqref{e:Ccurl4a} and \eqref{e:Ccurl4b} gives
\begin{multline}\label{e:Ccurl4c}
2\mathscr{C}(t)=4i\myIm c_3+\sum_{o_1\in\{-,+\}}o_1\Tr\left[\sigma_j^b\Pi_j^{o_1}\rho\Pi_j^{o_1}\right]\\
+\sum_{o_1\in\{-,+\}}o_1\Tr\left[\sigma_i^a(t)\sigma_j^b\sigma_i^a(t)\Pi_j^{o_1}\rho\Pi_j^{o_1}\right].
\end{multline}
Similarly, by applying the completeness relation $\Pi_j^{o_1}=\id_j-\Pi_j^{-o_1}$ and the corresponding spectral representation to the second occurrences of $\Pi_j^{o_1}$ in the first as well as the second trace of Eq.\ \eqref{e:Ccurl4c}, one arrives at
\begin{equation}\label{e:Ccurl5}
2\mathscr{C}(t)=2i\myIm\left(c_2+2c_3\right)+C(t)+1
\end{equation}
with
\begin{multline}\label{e:c2}
c_2=\sum_{o_1\in\{-,+\}}o_1 \Tr\left[\Pi_j^{-_b}\rho\Pi_j^{+_b}\sigma_i^a(t)\sigma_j^b\sigma_i^a(t)\right].
\end{multline}
The first term on the right-hand side of \eqref{e:Ccurl5} is purely imaginary, whereas $\mathscr{C}$ is real by definition \eqref{e:CorrProj}. Hence, by taking the real part on both sides of the Eq.\ \eqref{e:Ccurl5} one arrives at Eq.\ \eqref{e:CeqReC}.\\

\section{Proof of Eq.~\texorpdfstring{\eqref{e:CeqImC}}{(5)}}
\label{s:proof_rot}
Spin rotations at the lattice site $i$ around the $a$-axis by an angle $\theta$ are described by the unitary rotation operator
\begin{equation}
R_i^a(\theta) = \exp\left(-i\sigma_i^a \theta/2\right) \equiv \cos(\theta/2)-i\sin(\theta/2)\sigma_i^a.
\end{equation}
After step \ref{i:lasttime} of the rotation protocol the system is in the state
\begin{equation}
\mathscr{R}(t,\theta_1,\theta_2,\theta_3)\,\rho\,\mathscr{R}(t,\theta_1,\theta_2,\theta_3)^\dagger
\end{equation}
with
\begin{equation}
\mathscr{R}(t,\theta_1,\theta_2,\theta_3):=e^{-iHt}R_j^b(\theta_3)e^{iHt}R_i^a(\theta_2)e^{-iHt}R_j^b(\theta_1).
\end{equation}
Using this expression to calculate the expectation values of $\sigma_i^a$ on the left-hand side of Eq.\ \eqref{e:CeqImC} results in a lengthy expression with numerous combinations of sine and cosine terms. Simplifying these terms, either by hand in a tedious calculation or quickly using Mathematica, establishes the validity of Eq.\ \eqref{e:CeqImC}. This calculation makes repeated use of the involution $(\sigma_i^a)^2=\id$, which is an indication that the measurement protocol is unlike to hold beyond bivariate observables.

\bibliography{../../../MK,CG}

\end{document}